# Unravelling Ultralow Thermal Conductivity in Double Perovskite Cs$_2$AgBiBr$_6$: Dominant Wave-like Phonon Tunnelling, Strong Quartic Anharmonicity and Lattice Instability

Jiongzhi Zheng [1,2], Changpeng Lin [3,4], Chongjia Lin[1], Geoffroy Hautier[2], Ruiqiang Guo[5,†], Baoling Huang [1,6,7,‡]

[1]*Department of Mechanical and Aerospace Engineering, The Hong Kong University of Science and Technology, Clear Water Bay, Kowloon, Hong Kong*

[2]*Thayer School of Engineering, Dartmouth College, Hanover, New Hampshire, 03755, USA*

[3]*Theory and Simulation of Materials (THEOS), École Polytechnique Fédérale de Lausanne, CH-1015 Lausanne, Switzerland*

[4]*National Centre for Computational Design and Discovery of Novel Materials (MARVEL), École Polytechnique Fédérale de Lausanne, CH-1015 Lausanne, Switzerland*

[5]*Thermal Science Research Center, Shandong Institute of Advanced Technology, Jinan, Shandong Province, 250103, China*

[6]*HKUST Foshan Research Institute for Smart Manufacturing, Hong Kong University of Science and Technology, Clear Water Bay, Kowloon, Hong Kong, China*

[7]*HKUST Shenzhen-Hong Kong Collaborative Innovation Research Institute, Futian, Shenzhen 518055, China*

## Abstract

Efficient manipulation of thermal energy in halide perovskites is crucial for their opto-electronic, photovoltaic and thermoelectric applications. However, understanding the lattice dynamics and heat transport physics in the lead-free halide double perovskites remains an outstanding challenge due to their lattice dynamical instability and strong anharmonicity. In this work, we investigate the microscopic mechanisms of anharmonic lattice dynamics and thermal transport in lead-free halide double perovskite Cs$_2$AgBiBr$_6$ from first principles. We combine self-consistent phonon calculations with bubble diagram correction and a unified theory of lattice thermal transport that considers both the particle-like phonon propagation and wave-like tunnelling of phonons. An ultra-low thermal conductivity at room temperature (~0.21 Wm$^{-1}$K$^{-1}$) is predicted with weak temperature dependence (~$T^{0.34}$), in sharp contrast to the conventional ~$T^{-1}$ dependence. Particularly, the vibrational properties of Cs$_2$AgBiBr$_6$ are featured by strong anharmonicity and wave-like tunnelling of phonons. Anharmonic phonon renormalization from both the cubic and quartic anharmonicities are found essential in precisely predicting the phase transition temperature in Cs$_2$AgBiBr$_6$ while the negative phonon energy shifts induced by cubic anharmonicity has a significant influence on particle-like phonon propagation. Further, the contribution of the wave-like tunnelling to the total thermal conductivity surpasses that of the particle-like propagation above around 310 K, indicating the breakdown of the phonon gas picture conventionally used in the Peierls-Boltzmann Transport Equation. Importantly, further including four-phonon scatterings is required in achieving the dominance of wave-like tunnelling, as compared to the dominant particle-like propagation channel when considering only three-phonon scatterings. Our work highlights the importance of lattice anharmonicity and wave-like tunnelling of phonons in the thermal transport in lead-free halide double perovskites.

---

† ruiqiang.guo@iat.cn
‡ mebhuang@ust.hk

## I. INTRODUCTION

Halide perovskite materials, with a general chemical formula ABX$_3$, where A is a monovalent cation (Cs$^+$, methylammonium (CH$_3$NH$_3^+$) or formamidinium (CH(NH$_2$)$_2^+$)), B is a bivalent cation (Pb$^{2+}$ or Sn$^{2+}$), and X is a halide anion (Cl$^-$, Br$^-$ or I$^-$), have emerged as a new class of materials with outstanding optoelectronic properties. Generally, lead-based halide perovskites possess unique merits as optoelectronic materials, exhibiting long carrier lifetime and diffusion length, tuneable direct bandgap, high absorption coefficient and low trap density [1-3]. These properties make halide perovskites an ideal option for photovoltaic and optoelectronic devices in recent years. However, lead-based perovskite materials still encounter two crucial challenges, i.e., poor long-term stability and the toxicity of lead elements [3]. A promising strategy to tackle these issues is to replace the divalent Pb$^{2+}$/ Sn$^{2+}$ cation with one monovalent and one trivalent metal cation in the lead halide perovskites to form the so-called lead-free double perovskites $A_2B^+B^{3+}X_6$ (A = K, Rb, Cs; $B^+$ = Li, Na, K, Rb, Cs, Ag; $B^{3+}$ = Al, Ga, In, Sb, Bi, Sc, Y, and X = F, Cl, Br, I) [3,4]. The structural and functional diversity and the long-term stability of lead-free double perovskites render them attractive candidates for optoelectronic devices.

Apart from the optoelectronic properties, halide perovskites also attract research interest in lattice dynamics and thermal transport characteristics, which affect their thermodynamical stability [5], opto-electronic and photovoltaic performances [6]. Particularly, the effects of lattice vibrations on electronic and optoelectronic properties have been widely investigated to slow down the cooling of charge carriers in halide perovskites [7-9]. Besides, the experimentally reported ultra-low thermal conductivity $\kappa_L$ [10-15], high carrier mobilities [12,16,17] in conjunction with large Seebeck coefficient [16,17] make halide perovskites attractive for thermoelectric applications. A high $ZT$ of 1~2 in hybrid halide perovskite was predicted by *ab initio* calculations [18]. Importantly, the ultra-low $\kappa_L$ of halide perovskites was attributed to the low phonon group velocities and short phonon lifetimes [19] stemming from the complicated crystal structures and extremely strong lattice anharmonicity. Despite the progress in experimentally and theoretically observed thermal properties of halide perovskites [12-15,20], oxide and fluoride perovskites [21], an accurate quantitative prediction of their lattice dynamical properties and a comprehensive microscopic understanding of their thermal transport properties are still in infancy.

The outstanding challenges in theoretical understanding of phonon-related properties in halide perovskites can be traced back to the strong lattice anharmonicity, which makes the conventional approaches within the harmonic approximation framework fail [22-25]. To overcome the shortcoming of harmonic approximation, the anharmonic phonon renormalization techniques [22-24,26] have been applied to predict the phase transition temperature and thermal properties of halide perovskites at finite temperatures [27-29]. However, most of works consider only cubic-anharmonicity and the explicit treatment of effects of both cubic- and quartic-anharmonicities on phonon energies and transport channels remains challenging. Also, the particle-like phonon picture in describing thermal transport physics generally breaks down in highly anharmonic materials when the phonon linewidths are larger than the interbranch spacings [30-33]. For example, using the conventional first-principles-based Peierls-Boltzmann transport equation (PBTE) method [34], Lee *et al.* [12] failed to reproduce the experimentally measured ultra-low $\kappa_L$ in all-inorganic halide perovskites. Going beyond the conventional particle-like description of phonons in PBTE, Simoncelli *et al.* [30] explained the experimentally observed lattice thermal conductivity in the complex perovskite CsPbBr$_3$ by further considering wave-like tunnelling of phonons. The unified theory of thermal transport adopted in the above literature [30] can accurately reproduce the $\kappa_L$; however, it considers only the three-phonon scattering processes within the harmonic approximation treatment (The error cancellation due to phonon renormalization and four phonon scattering rates may occurs in the $\kappa_L$ of CsPbBr$_3$, similar to that observed in PbTe [35]). Therefore, a comprehensive understanding of the lattice dynamics and thermal properties in lead-free halide double perovskites have been rarely reported.

In this paper, we systematically investigate the anharmonic lattice dynamics and thermal transport properties in a benchmark lead-free halide double perovskite Cs$_2$AgBiBr$_6$ at the atomic level. We have applied a state-of-the-art unified theory of thermal transport to predict the lattice thermal conductivity $\kappa_L$ in perovskite Cs$_2$AgBiBr$_6$ by considering the isotope-, three- (3ph) and four-phonon (4ph) scatterings, anharmonic phonon renormalization from both cubic and quartic anharmonicities, particle-like propagation and wavelike tunnelling transport channels. Using the anharmonic phonon renormalization technique, we find that the soft modes are associated with the tilting of the AgBr$_6$ and BiBr$_6$ octahedra units and become hardened with increasing temperature. The cubic-to-tetragonal phase transition temperature of $Cs_2AgBiBr_6$ is predicted as ~119-138 K, agreeing well with the experimental value. Using the unified theory of thermal transport, we show that both the 4ph scattering processes and wavelike tunnelling of phonons play a crucial role in predicting the $\kappa_L$ of $Cs_2AgBiBr_6$. Particularly, considering only the 3ph scattering, the populations' conductivity accounts for more than 50% of the total $\kappa_L$ over the entire temperature range. By further considering 4ph scatterings, the coherences' conductivity ascribing from the wave-like tunnelling channel dominates the total $\kappa_L$ above ~310 K. This study highlights the importance of wave-like tunnelling and quartic anharmonicity in describing heat conduction in $Cs_2AgBiBr_6$.

## II. METHODOLOGY AND COMPUTATIONAL DETAILS
### a. First-principles simulations

In this work, the *ab-initio* calculations within the framework of density functional theory (DFT) [36] using the Vienna *Ab-initio* Simulation Package (VASP) [37] were performed for structural optimization of $Cs_2AgBiBr_6$ crystal, wherein the projector-augmented wave (PAW) [38] method was used to treat the $Cs(5s^25p^66s^1)$, $Ag(4d^{10}5s^1)$, $Bi(5d^{10}6s^26p^3)$ and $Br(4s^24p^5)$ shells as valence states. The revised Perdew-Burke-Ernzerhof (PBE) version for solids, i.e., PBEsol [39], of the generalized gradient approximation (GGA) [40] was used for the exchange-correlation functional. The ionic positions and unit cell geometry were fully optimized using a plane-wave cutoff energy of 800 eV and a 10×10×10 Monkhorst-Pack electronic *k*-point mesh, with a tight force convergence criterion of $10^{-5}$ eV·Å$^{-1}$ and a tight energy convergence criterion of $10^{-8}$ eV. The resulting relaxed lattice constant is $a$ = b = c = 5.602 Å, in good agreement with the experiment ($a$=5.639 Å) [41,42] for $Cs_2AgBiBr_6$ crystal with a $Fm\overline{3}m$ space group. Considering that the long-term dipole-dipole interaction in the polar $Cs_2AgBiBr_6$ crystal will lead to the longitudinal (LO) and transverse optical (TO) phonon modes splitting near the Brillouin zone centre, the non-analytic corrections are further included in the dynamical matrix and the dielectric tensor $\varepsilon$ and the Born effective charges $Z$ of $Cs_2AgBiBr_6$ are calculated using the density functional perturbation theory (DFPT) [43]. The corresponding parameters are computed to be $\varepsilon^\infty = 5.392$, $Z^*(Cs) = 1.356$, $Z^*(Ag) = 1.795$, $Z^*(Bi) = 4.805$, $Z^*(Br)_\perp = -0.769$, and $Z^*(Br)_\text{II} = -3.114$, which are consistent with previous theorectical calculations [29]. Given the experimental thermal expansion of $Cs_2AgBiBr_6$ in the temperature range of 100 – 300 K is relatively small (~0.53% -- from 11.21 to 11.27 Å) [41], we have opted not to consider the thermal expansion in the current calculations [For more details, please refer to Fig. S1 in the SM [44]].

The 0-K harmonic interatomic force constants (IFCs) calculation was performed using the finite-displacement approach [45], with a 2×2×2 supercell and 4×4×4 Monkhorst-Pack electronic $k$-point mesh, and a plane-wave cutoff energy of 800 eV in VASP. Additionally, the same supercell dimensions and DFT calculation settings were used to map out the potential energy surfaces (PESs). To extract anharmonic IFCs efficiently, the machine-learning-based method called the compressive sensing lattice dynamics method (CSLD) was used [46,47], where the compressive technique [48] was implemented to collect the physically important terms of the anharmonic IFCs using the limited displacement-force datasets [22]. Within the framework of the CSLD method, the *ab initio* molecular dynamics (AIMD) simulation was performed with a 2×2×2 supercell to generate the random atomic displacements, with a 2×2×2 Monkhorst-Pack electronic $k$-point mesh and an energy convergence criterion of $10^{-5}$ eV. The AIMD simulation of the NVT ensemble with the Nosé Hoover thermostat at 500 K was ran for 3000-time steps using a 2-fs time step. By disregarding the first 500 steps from the obtained AIMD trajectories, 100 atomic structures were sampled at equal intervals. After that, a random-direction displacement of 0.1 Å for all atoms was imposed to reduce the cross-correlations between the atomic structures. Furthermore, accurate DFT calculations were performed to obtain the training and cross-validation data, on the basis of the sampled atomic structures, with 4×4×4 Monkhorst-Pack electronic *k*-point mesh and energy convergence better than $10^{-8}$ eV. Finally, using the displacement-force datasets and 0-K harmonic IFCs obtained, the least absolute shrinkage and selection operator (LASSO) technique [49] was used to extract the anharmonic IFCs up to the sixth order. The real-space cutoff radii of 8.47 Å, 6.35 Å, 4.23 Å and 3.18 Å were applied for the cubic, quartic, quintic, and sextic IFCs extraction, respectively. We observe a fitting error of 6.2% for higher-order IFCs in the context of atomic force. It is noteworthy that this error is exceptionally small, mirroring values reported in the existing literatures [22,50,51]. In this work, the IFCs estimation was performed by using the **ALAMODE** package [22,52,53].

### b. Anharmonic phonon renormalization

The anharmonic phonon energy renormalization was performed using the self-consistent phonon approximation (SCP) formulated in the reciprocal space [26,53], as implemented in the **ALAMODE** package [22,53]. Considering only the first-order correction from quartic anharmonicity, i.e., the loop diagram, the resulting SCP equation in the diagonal form neglecting polarization mixing (PM) can be written as

$$\Omega_q^2 = \omega_q^2 + 2\Omega_q I_q, \tag{1}$$

where $\omega_q$ is the bare harmonic phonon frequency associated with the phonon mode $q$, and $\Omega_q$ is the anharmonically renormalized phonon frequency at finite temperatures. The quantity $I_q$ can be defined as

$$I_q = \frac{1}{8N}\sum_{q'} \frac{\hbar V^{(4)}(q;-q;q';-q')}{4\Omega_q \Omega_{q'}}[1 + 2n(\Omega_{q'})], \tag{2}$$

where $N$, $\hbar$, $n$ and $V^{(4)}(q;-q;q';-q')$ are the total number of sampled phonon wavevectors in the first Brillouin zone, the reduced Planck constant, the Bose-Einstein distribution and the reciprocal representation of 4$^{\text{th}}$-order IFCs, respectively. Note that the off-diagonal terms of the phonon loop self-energy for PM [22] were found to be crucial for some crystal systems [54,55], thus, we also further included it in the anharmonic phonon energy renormalization calculations. In this work, the $q$-mesh of the SCP calculations was set to 2×2×2, which is equivalent to a 2×2×2 supercell adopted by the real-space-based phonon energy renormalization technique. Meanwhile, the SCP $q$-mesh of 2×2×2 was used to correspond to the supercell dimensions of the second-order IFCs extracted in the real space.

Using the above first-order SCP technique that only obtains the renormalized phonon energies resulting from the quartic anharmonicity, i.e., the loop diagram, the anharmonic phonon energies of highly anharmonic compounds may be overestimated [56,57]. On top of the renormalized phonon energies calculated by the first-order SCP, the renormalized phonon energies arising from the cubic anharmonicity, i.e., the dominant bubble diagram, can be estimated using the following self-consistent equation within quasi-particle approximation [57]

$$\left(\Omega_q^B\right)^2 = \Omega_q^2 - 2\Omega_q \mathrm{Re} \sum_q^B [G, \Phi_3](\Omega=\Omega_q^B), \qquad (3)$$

where $\sum_q^B[G, \Phi_3](\Omega_q)$ is the phonon frequency-dependent bubble self-energy, $B$ denotes the bubble diagram, $\Phi_3$ is the third-order force constant explicitly included in the anharmonic self-energy calculation. Note that several treatments within quasiparticle (QP) approximation including QP[0], QP[S] and QP-NL can be employed to solve Eq. 3 and obtain the fully renormalized phonon energies due to cubic and quartic anharmonicities [57]. The QP-NL treatment was used in this work due to its reliable predictions. Additionally, the extra phonon energy shifts arising from the tadpole diagram due to the cubic anharmonicity were found to be small and negligible, which was also reported for many crystals [54-56]. Note that the SCP and SCPB were denoted as the shorthand notation of the first-order SCP calculation and SCP calculation with bubble diagram correction in this work.

### c. Phonon scattering rates and unified thermal transport theory

With the renormalized harmonic (2nd-order) and original anharmonic IFCs (3rd and 4th-order), the phonon scattering rates resulting from phonon-phonon and phonon-isotope interaction processes, as the key ingredients entering the thermal transport equation [34], can be calculated using Fermi's golden rule of perturbation theory [58]. In this study, the intrinsic multi-phonon interaction mainly includes three-phonon (3ph) and four-phonon (4ph) scattering processes and the corresponding expressions within single-mode relaxation time approximation (SMRTA) can be written as [25,56,58]

$$\Gamma_q^{3ph} = \sum_{q'q''} \left\{ \frac{1}{2}\left(1 + n_{q'}^0 + n_{q''}^0\right)\zeta_- + \left(n_{q'}^0 - n_{q''}^0\right)\zeta_+ \right\}, \quad (4)$$

$$\Gamma_q^{4ph} = \sum_{q'q''q'''} \left\{ \frac{1}{6}\frac{n_{q'}^0 n_{q''}^0 n_{q'''}^0}{n_q^0}\zeta_{--} + \frac{1}{2}\frac{\left(1+n_{q'}^0\right)n_{q''}^0 n_{q'''}^0}{n_q^0}\zeta_{+-} + \frac{1}{2}\frac{\left(1+n_{q'}^0\right)\left(1+n_{q''}^0\right)n_{q'''}^0}{n_q^0}\zeta_{++} \right\}, \quad (5)$$

where $\zeta_\pm$ and $\zeta_{\pm\pm}$ are defined as

$$\zeta_\pm = \frac{\pi\hbar}{4N}\left|V^{(3)}\left(q,\pm q',-q''\right)\right|^2 \Delta_\pm \frac{\delta\left(\Omega_q \pm \Omega_{q'} - \Omega_{q''}\right)}{\Omega_q \Omega_{q'} \Omega_{q''}}, \quad (6)$$

$$\zeta_{\pm\pm} = \frac{\pi\hbar^2}{8N^2}\left|V^{(4)}\left(q,\pm q',\pm q'',-q'''\right)\right|^2 \Delta_{\pm\pm} \frac{\delta\left(\Omega_q \pm \Omega_{q'} \pm \Omega_{q''} - \Omega_{q'''}\right)}{\Omega_q \Omega_{q'} \Omega_{q''} \Omega_{q'''}}, \quad (7)$$

where $\Gamma_q^{3ph}$ and $\Gamma_q^{4ph}$ denote the 3ph and 4ph scattering rates, respectively, the phonon mode $q$ is a shorthand notation for a composite index of the wavevector **q** and phonon branch $j$, $V^{(3)}\left(q,\pm q',-q''\right)$ and $V^{(4)}\left(q,\pm q',\pm q'',-q'''\right)$ are the reciprocal representation of 3rd- and 4th-order IFCs, respectively [59], the delta function $\delta(\Omega)$ describes the selection rules for energy in both 3ph and 4ph scattering processes, the Kronecker deltas $\Delta_\pm$ and $\Delta_{\pm\pm}$ are shorthand notation for $\Delta_{q\pm q'-q'',Q}$ and $\Delta_{q\pm q'\pm q''-q''',Q}$, respectively, and account for the selection rules for momentum, where the reciprocal lattice vector $Q = 0$ indicates the normal (N) process and $Q \neq 0$ the Umklapp (U) process.

The phonon scattering term resulting from isotope can be written as [60]

$$\Gamma_q^{\text{isotope}} = \frac{\pi \Omega_q^2}{2N} \sum_{i \in u.c.} g(i) \left| e_q^*(i) \cdot e_{q'}(i) \right|^2 \delta(\Omega - \Omega'), \tag{8}$$

where the mass variance, $g(i)$, is defined as $g(i) = \sum_s f_s(i)[1 - M_s(i)/\bar{M}(i)]^2 = \sum_s f_s(i)[\Delta M_s(i)/\bar{M}(i)]^2$, where $f_s(i)$ and $M_s(i)$ are the concentration and mass of the $sth$ isotope of atom $i$, respectively. $\bar{M}(i)$ denotes the average mass of the $ith$ atom in the primitive cell. $e_q(i)$ denotes the eigenfunction of phonon mode $q$ at the atom $i$.

Using Matthiessen's rule, the total phonon scattering rate $\Gamma_q$ can be expressed as

$$\Gamma_q = \Gamma_q^{\text{3ph}} + \Gamma_q^{\text{4ph}} + \Gamma_q^{\text{isotope}}, \tag{9}$$

The lattice thermal conductivity $\kappa_L$ can be calculated by a unified theory of thermal transport incorporating populations' and coherences' conductivities/contributions [30], and the detailed formula for $\kappa_L$ under the SMRTA is given as follows [30,56,61]:

$$\begin{aligned} \kappa_L^{P/C} = \frac{\hbar^2}{k_B T^2 V N} \sum_q \sum_{j,j'} \frac{\Omega_{qj} + \Omega_{qj'}}{2} v_{qjj'} \otimes v_{qj'j} \\ \cdot \frac{\Omega_{qj} n_{qj}(n_{qj}+1) + \Omega_{qj'} n_{qj'}(n_{qj'}+1)}{4(\Omega_{qj} - \Omega_{qj'})^2 + (\Gamma_{qj} + \Gamma_{qj'})^2} (\Gamma_{qj} + \Gamma_{qj'}) \end{aligned}, \tag{10}$$

where the superscripts $P$ and $C$ denote the populations' and coherences' contributions, respectively, $k_B$ is the Boltzmann constant, $T$ is the temperature, $V$ is the unit-cell volume and $v$ is the group velocity matix including both the intra- and inter-branch terms [62]. In equation (10), when $j = j'$, it calculates the populations' contribution (PBTE calculation result) $(\kappa_L^P)$, i.e., particle-like phonon propagation described by the diagonal terms of heat flux operators, otherwise, it gives the coherences' contribution ($\kappa_L^C$), i.e., wave-like tunelling of phonons described by the off-diagonal terms of heat flux operators (The terminology,i.e., diffuson, is also used to describe this operator In galsses). The total lattice thermal conductivity $\kappa_L$ can be obtained by summing over $\kappa_L^P$ and $\kappa_L^C$, namely $\kappa_L = \kappa_L^P + \kappa_L^C$. In this work, the $\boldsymbol{q}$ mesh for 3ph scattering processes was set to 12×12×12, which gives well-converged results for the crystalline $Cs_2AgBiBr_6$. Considering the huge computational cost, a $\boldsymbol{q}$ mesh of 8×8×8 with a scalebroad parameter of 0.06 were used for 4ph scattering processes. Note that the iterative scheme to PBTE is only applied for 3ph scattering processes in this work, while the 4ph scattering processes are treated at the SMRTA level considering the exceptionally large memory demands [56,59]. Thermal transport calculations including particle-like propagation and wave-like tunnelling transport channels were performed by using the **ShengBTE** [63] and **FourPhonon** [59] packages, as well as our in-house code [54].

## III. RESULTS AND DISCUSSION
### a) Lattice dynamics and anharmonicity

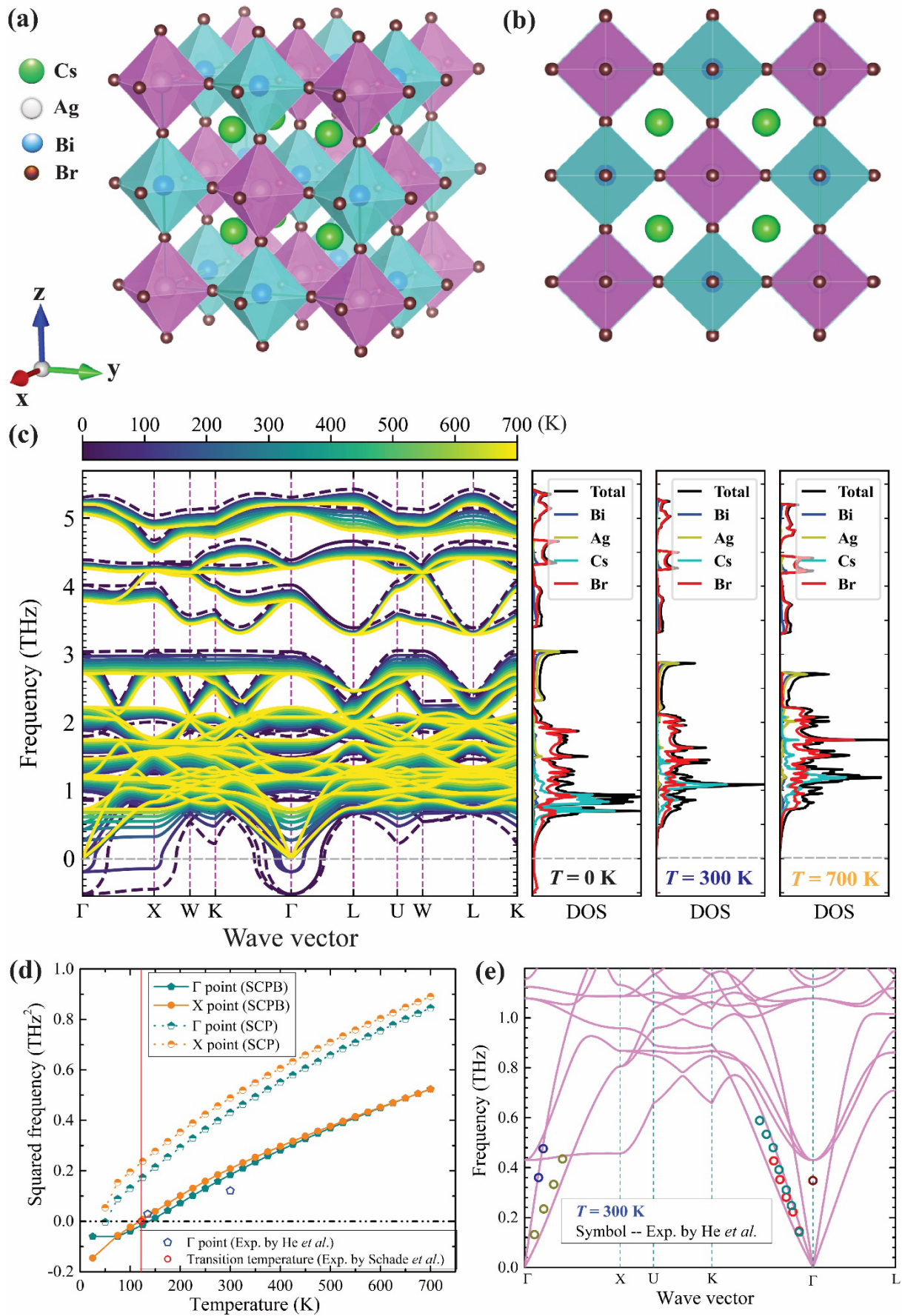

FIG. 1. (a) Crystal structure of the cubic double perovskite $Cs_2AgBiBr_6$, which features two different sub-lattices formed by $AgBr_6$ and $BiBr_6$ octahedra units, respectively, together with Cs atoms. The solid green lines inside the sub-lattice depict a conventional unit cell, and the green, white, blue, and brown spheres represent Cs, Ag, Bi and Br atoms, respectively. (b) View of the *yz* plane of the crystal structure. (c) Calculated anharmonically renormalized phonon dispersions at finite temperatures (from $T$ = 100 to 700 K) compared with the phonon dispersions computed by the harmonic approximation treatment at $T$ = 0 K. The calculated atom-decomposed partial and total phonon densities of states at $T$ = 0, 300, and 700 K are shown on the right panel, respectively. (d) Calculated temperature-dependent squared phonon frequency of soft modes at the Γ and X points using the SCP and SCPB approximations, respectively. The empty red circle indicates the experimentally measured phase transition temperature (~122 K) [41,64,65], and the empty blue pentagon indicates the experimental phonon energies of the soft mode at the Γ point at 135 and 300 K, respectively [65]. (e) Calculated temperature-dependent renormalized phonon dispersions along the high-symmetry paths compared with experimental results at 300 K [65].

We start by investigating the vibrational properties of double perovskite $Cs_2AgBiBr_6$ in its cubic phase (as shown in Figs. 1(a-b)) using the static harmonic approximation (HA) approach. The calculated harmonic phonon dispersions (represented by dark blue dash lines) and density of states (DOS) of $Cs_2AgBiBr_6$ are shown in Fig. 1(c). One of the most notable characteristics of the harmonic phonon dispersions in cubic $Cs_2AgBiBr_6$ is the occurrence of soft modes with negative phonon energies, which are observed at the Γ and X points in the Brillouin zone. This feature suggests the presence of dynamical instability at low temperatures [see in Fig. 1(c)]. Upon analysing the vibrational properties of $Cs_2AgBiBr_6$, we have observed that the soft modes at the Γ and X points are associated with the tilting of the $AgBr_6$ and $BiBr_6$ octahedra units, respectively, which was also reported by Klarbring *et al.* [29]. By applying a comprehensive anharmonic phonon renormalization technique, these soft modes are found to become significantly hardened with increasing temperature [see anharmonically renormalized phonon dispersions and DOS in Fig. 1(c)]. The soft modes in $Cs_2AgBiBr_6$ are primarily contributed by Br atoms, which is evident from the atom-decomposed partial DOS [see the right panel in Fig. 1(c)] and atomic participation ratio of Br atoms projected onto the phonon bands [see Fig. S2 in supplemental material (SM) [44]]. Apart from the Br-dominated soft modes, the low-frequency phonon modes (≤ 2.2 THz), contributed by Cs atoms, also undergo a gradual stiffening [see Figs. 1(c) and S2(b) in the SM [44]]. The temperature-dependent stiffening behaviour of low-frequency modes is attributed to the large atomic displacements that arise from the relatively weak bonding interactions [see Fig. S3 in the SM [44]]. In contrast, the high-frequency phonon modes (> 2.2 THz) in Cs$_2$AgBiBr$_6$ are predominantly contributed by Ag and Br atoms [see Figs. S2(a) and (d) in the SM [44]] and exhibits a temperature-dependent softening, consistent with experimental observations [64].

It is worth noting that in some crystals, the phonon energy shifts resulting from 3ph interaction processes, specifically the bubble diagram (including those in the SCPB approximation), can be ignored due to their minor contributions [53] and computational complexity [57]. In the case of crystalline Cs$_2$AgBiBr$_6$, considering only 4ph processes, i.e., the loop diagram (SCP approximation), in anharmonic phonon renormalization shows only phonon stiffening [see Fig. S4 in the SM [44]]. Additionally, the anharmonically renormalized phonon energies in Cs$_2$AgBiBr$_6$ are stabilized above 50 K when considering only 4ph processes [see Fig. 1(d)]. This value is close to the prediction of Klarbring *et al.* [29] but significantly lower than the experimental value of approximately 122 K [41,64,65]. The observed discrepancy in the phase transition temperature of Cs$_2$AgBiBr$_6$, between the value computed solely from 4ph processes and the experimental result, underscores the crucial role of the negative phonon energy shifts from 3ph processes in accurately determining phonon energies. By incorporating the 3ph processes in phonon renormalization calculations, it predicts that the cubic-to-tetragonal phase transition of Cs$_2$AgBiBr$_6$ occurs at ~119 K (phonons collapse at X point) and ~138 K (phonons collapse at Γ point) [see in Fig. 1(d)], in good agreement with experimental measurements [41,64,65]. Furthermore, in comparison to the SCP approximation, the SCPB approximation provides more accurate predictions for the phonon energies of the soft mode at Γ point, as observed in experiments conducted at 135 and 300 K [see in Fig. 1(d)] [65]. To validate out calculations, we compared the phonon dispersions along high-symmetry paths calculated using the SCPB approximation with the experimental results obtained from inelastic neutron scattering measurements [65]. The comparison shows good agreement between the calculated and experimental values [see Fig. 1(e) and Figs. S5 and S6 in SM [44]]. The importance of both 3ph and 4ph processes in reproducing the experimental lattice vibrational properties of $Cs_2AgBiBr_6$ crystal is highlighted by these results, as has been demonstrated for several other compounds, e.g., $BaZrO_3$, $CsPbBr_3$ and SnSe [54,57,66].

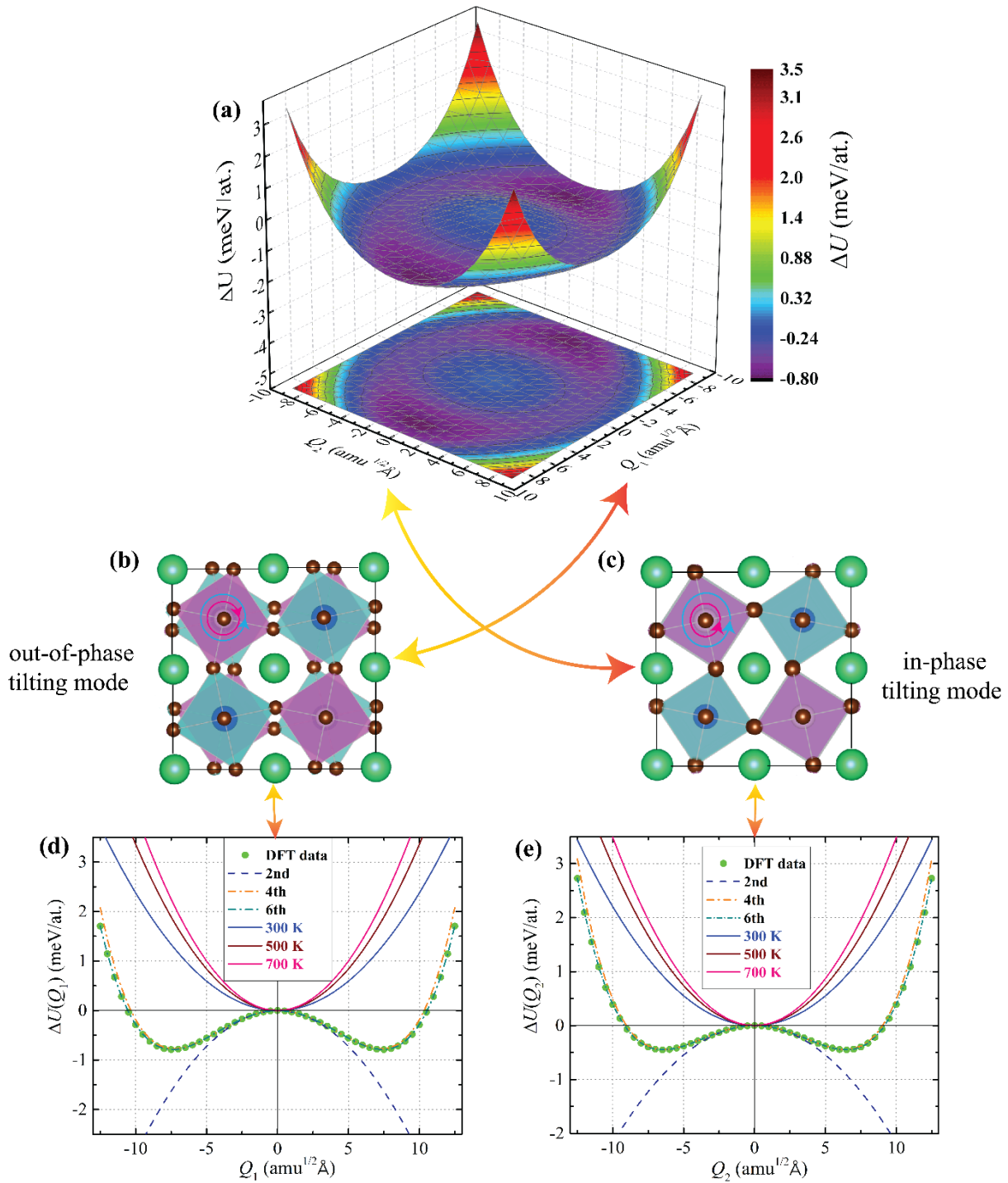

FIG. 2. (a) Calculated two-dimensional (2D) potential energy surface of $Cs_2AgBiBr_6$ as a function of normal mode coordinates $Q_1$ and $Q_2$. Here the soft modes at the Γ and X points associated with the out-of-phase and in-phase

tilts are selected to generate the corresponding configurations. (b) Crystal structure of $Cs_2AgBiBr_6$ with displacement pattern of the soft mode at the Γ point, wherein the red and light blue circles with arrows are depicted for the displacements of $AgBr_6$ and $BiBr_6$ octahedron units (out-of-phase tilting), respectively. (c) The same as (b), but with displacement pattern of the soft mode at the X point and (in-phase tilting). (d) DFT-calculated double well potential energy surface (solid green disks) of the soft mode at the Γ point as a function of the vibrational amplitude ($Q_1$) in the normal mode coordinate. The blue, yellow and green dash lines show the potential energy surface decomposed to the second, fourth and sixth orders, respectively. The blue, brown and pink solid lines represent the renormalized harmonic potentials at $T$ = 300, 500, and 700 K, respectively. (e) The same as (d), but for soft mode at the X point and the normal mode coordinate ($Q_2$).

To gain insight into lattice instability, anharmonicity, and phonon energy shift in $Cs_2AgBiBr_6$ crystal, we mapped out the 2D potential energy surface (PES) [67] of soft modes at the Γ and X points. The PES was plotted along two corresponding normal-mode coordinates ($Q_1$, and $Q_2$) [see Fig. 2(a)]. The soft modes at the Γ and X points are associated with the out-of-phase and in-phase tilting of the $AgBr_6$ and $BiBr_6$ octahedra units, respectively, as shown in Figs. 2(b-c). As anticipated, the minimum energy is located outside the zero-tilt amplitude ($Q_1 = Q_2 = 0$) of both soft modes at the Γ and X points, indicating lattice instability or dynamical instability [see Fig. 2(a)] [67]. The 2D PES mapping in Fig. 2(a) reveals that the minimum energy of the soft mode at Γ point is lower than that of soft mode at the X point. This observation suggests the dominant role of the out-of-phase tilting mode, associated with the $AgBr_6$ octahedra, in the cubic-to-tetragonal phase transition of $Cs_2AgBiBr_6$. Although the minimum energies exist outside the zero-tilt amplitude for both soft modes, the corresponding minima are relatively shallow, with values of about -0.79 and -0.45 meV/atom, respectively. This observation suggests that the high-temperature cubic phase of $Cs_2AgBiBr_6$ crystal is effectively an average structure at relatively low temperatures, similar to the specific phase of some previously reported compounds such as $CsSnI_3$, $CsSnBr_3$ and $CsSnCl_3$ [68-70].

To gain a more intuitive understanding of the lattice anharmonicity of $Cs_2AgBiBr_6$, we projected the 2D PES onto 1D PES for both the soft modes at Γ and X points [see Fig. 2(d-e)]. As expected, each soft mode forms a double-well potential, suggesting strong lattice anharmonicity [56]. Indeed, the anharmonic double-well potentials of soft modes can be decomposed up to the sixth order, although the contribution of the sextic terms is minor [see

Figs. 2(d-e)]. Additionally, the harmonic potential (blue dash line) of soft modes has a negative coefficient, which is responsible for the imaginary frequency [see Fig. 1(b)], and therefore leads to the failure of the conventional harmonic approximation (HA) treatment [22,34,56]. It is worth mentioning that although the PESs of the soft modes exhibit anharmonic behaviour up to the sixth-order terms, the contribution of these higher-order terms is minor. Therefore, in our anharmonic phonon renormalization calculations, we only consider the anharmonic contributions up to fourth-order terms. The lattice anharmonicity was utilized in the anharmonic phonon renormalization to gradually transform the negative harmonic potentials of the soft modes into positive ones at finite temperatures, such as 300 K, 500 K and 700 K [see Figs. 2(d-e)]. These positive potentials represent the effective harmonic potential with temperature effects taken into account.

### b) Lattice thermal conductivity

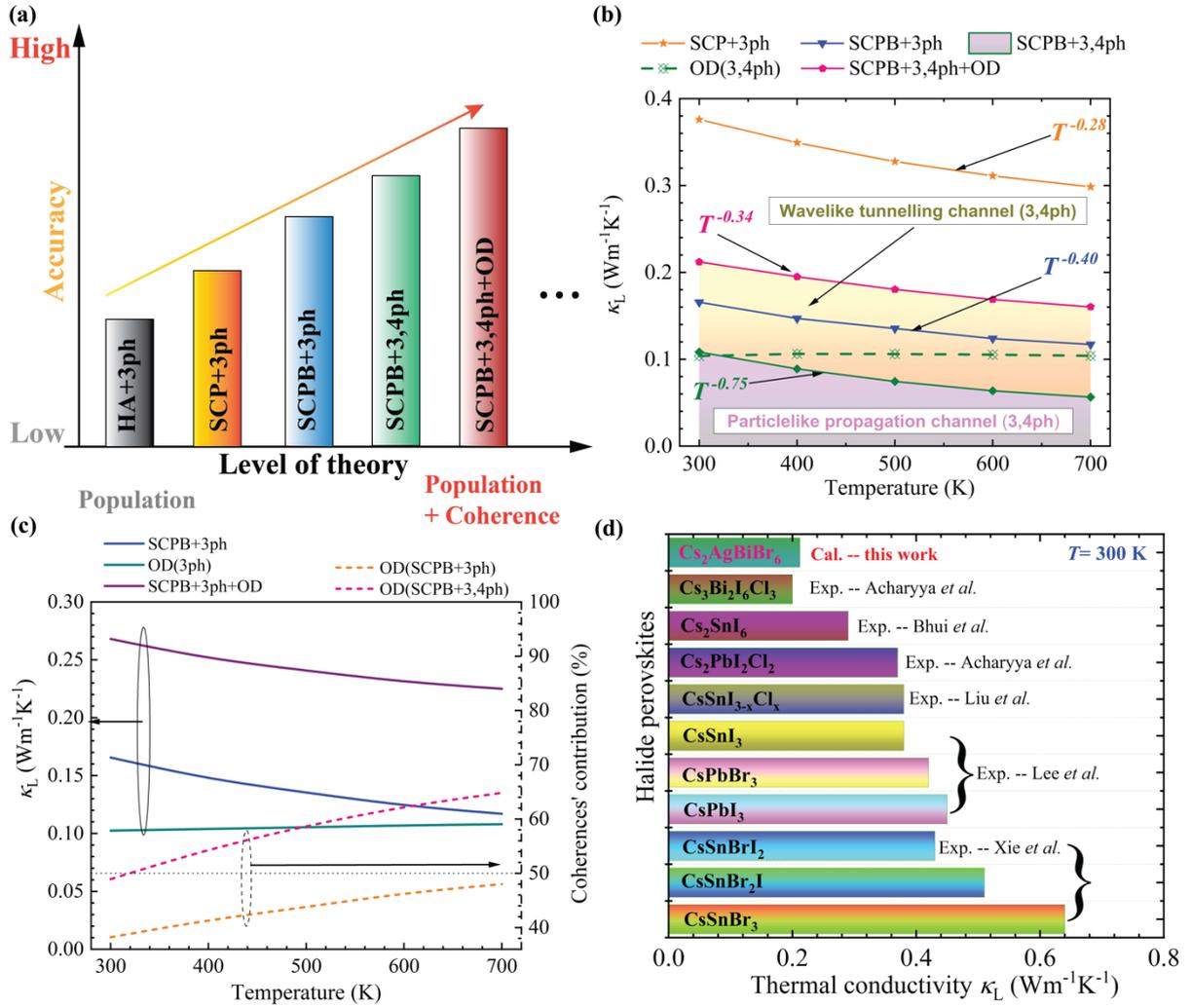

FIG. 3. (a) A schematic diagram depicting the prediction accuracy of lattice thermal conductivity using various hierarchical thermal transport models, namely, the HA+3ph, SCP+3ph, SCPB+3ph, SCPB+3,4ph, and SCPB+3,4ph+OD models, respectively. The ellipsis indicates the theory of thermal transport beyond the SCPB+3,4ph+OD model and it may further include anharmonic heat flux operators [71]. (b) Calculated temperature-dependent lattice thermal conductivity $\kappa_L$ using different levels of thermal transport theory including the SCP/SCPB + 3ph, SCPB + 3,4ph + OD models. The yellow and pink shaded areas show the contributions from phonon and coherence thermal transport channels calculated using the SCPB+3,4ph+OD model, respectively. (c) Temperature-dependent populations', coherences' and total thermal conductivity calculated using SCPB+3ph+OD model and the coherences' contributions, percentage wise, to total thermal conductivity calculated using SCPB+3/3,4ph models. (d) Comparison of our predicted lattice thermal conductivity of crystalline $Cs_2AgBiBr_6$ with other experimentally measured ultra-low thermal conductivity of all-inorganic halide perovskites at room temperature [12-15,72,73].

Next, we investigate the effects of phonon energy renormalization and multi-phonon interaction from both cubic and quartic anharmonicities on lattice thermal conductivity $\kappa_L$. To achieve this, we employ a hierarchy of theoretical approaches with increasing accuracy, as depicted in Fig. 3(a). Due to the failure of conventional harmonic approximation (HA) treatment in capturing the lattice dynamics of $Cs_2AgBiBr_6$, we begin our analysis by examining the $\kappa_L$ obtained using the SCP+3ph model. The populations' thermal conductivity $\kappa_L^P$ was computed using the SCP+3ph model, which takes into account renormalized phonon energies from only quartic anharmonicity and employs only 3ph scattering processes. The resulting value at 300 K is ~ 0.38 Wm$^{-1}$K$^{-1}$, as shown in Fig. 3(b). This predicted $\kappa_L^P$ is consistent with the result reported by Klarbring *et al.* [29], although their method utilized the temperature-dependent effective potential approach based on AIMD simulations. As discussed earlier, the anharmonic phonon renormalization using only quartic anharmonicity can lead to an overestimation of the phonon energies and phase transition temperature of Cs$_2$AgBiBr$_6$. To account for extra phonon energy shifts from cubic anharmonicity, we move up the theory level progressively to the SCPB+3ph model. As a result, we observe a remarkable decrease in $\kappa_L^P$ from ~ 0.38 to ~ 0.17 Wm$^{-1}$K$^{-1}$ (~55.95% reduction) at 300 K, as demonstrated in Fig. 3(b). This observation highlights the significant influence of cubic anharmonicity not only on lattice dynamics but also on phonon transport in crystalline Cs$_2$AgBiBr$_6$. A similar phenomenon has been reported for simple cubic perovskites, e.g., CsSnBr$_3$, BaZrO$_3$ and CsPbBr$_3$, in previous studies [28,54,57]. To properly account for the anharmonicity of highly anharmonic compounds, it is necessary to include multi-phonon scatterings resulting from quartic anharmonicity [25,56]. By further incorporating 4ph scattering processes into the SCPB+3ph model, the improved SCPB+3,4ph model predicts a further reduction in $\kappa_L^P$. Specifically, the predicted $\kappa_L^P$ decreases to ~0.11 and ~0.06 Wm$^{-1}$K$^{-1}$ (~34.6 and 51.8% reduction) at 300 and 700 K, respectively [see Fig. 3(b)]. This highlights the crucial role of 4ph scattering processes in determining the $\kappa_L^P$ of highly anharmonic Cs$_2$AgBiBr$_6$. Recent studies have emphasized the importance of coherences' contributions from non-diagonal terms of heat flux operators, i.e., wave-like tunnelling channel, in compounds with ultra-low $\kappa_L$ or/and strong anharmonicity [30,31,56,61]. By incorporating the wave-like tunnelling transport channel [30,61] through non-diagonal terms of heat flux operators, beyond the phonon-gas model, the SCPB+3,4ph+OD model results in a substantial increase in $\kappa_L$ to 0.21 and 0.16 Wm$^{-1}$K$^{-1}$ (~93.9 and 184.2% enhancement) at 300 and 700 K, respectively. Interestingly, using the SCPB+3,4ph+OD model reveals that the total $\kappa_L$ contributed from the wavelike tunnelling channel comprises more than 50% of the total $\kappa_L$ above 340 K [see Figs. 3(b-c)]. In contrast, when considering only 3ph scatterings, the phonon propagation channel dominates the total $\kappa_L$, accounting for more than 50% of the total heat conduction over the entire temperature range [see Fig.3(c)]. The presence of strong 4ph scatterings triggers the inter-conversion of the dominant role of heat conduction between the particle-like phonon propagation and the wave-like tunnelling channels in Cs$_2$AgBiBr$_6$. This observation emphasizes the limitations of the conventional phonon-gas model and highlights the importance of considering higher-order anharmonicity and coherences' effects in advanced thermal transport models to accurately describe the thermal transport of Cs$_2$AgBiBr$_6$.

The dominant role of coherences' contributions $\kappa_L^C$ to the total $\kappa_L$ over a wide range of temperatures implies that the heat transfer physics in Cs$_2$AgBiBr$_6$ exhibits similar features to those observed in glasses. We proceed to analyze the temperature dependence of the lattice thermal conductivity of Cs$_2$AgBiBr$_6$ calculated using different levels of theory. The predicted $\kappa_L^P$ predicted by the SCP+3ph model follows an extremely weak temperature dependence of $\sim T^{-0.28}$ in the temperature range of 300-700 K [see Fig. 3(b)]. This observation challenges the conventional understanding that the $\kappa_L^P$ computed by the lowest-order perturbation theory [74] follows a temperature dependence of $\sim T^{-1}$. Using the SCPB+3ph model, the temperature dependence of the $\kappa_L^P$ becomes stronger and follows $\sim T^{-0.40}$, which is a result of the softening of phonon frequency. Further including 4ph scattering processes leads to a significant change in the temperature dependence of the $\kappa_L^P$, which is remarkably enhanced to $\sim T^{-0.75}$. This observation is consistent with the findings of Feng *et al.* [58] and Zheng *et al.* [54]. By utilizing the SCPB+3,4ph+OD model, the $\kappa_L$ exhibits a weak temperature dependence of $\sim T^{-0.34}$, resembling the experimentally observed out-of-plane temperature-dependent behaviour of $\kappa_L$ in $Cs_2PbI_2Cl_2$ [14] [see Fig. 3(b)].

To validate our computational results, we compared the calculated $\kappa_L$ of $Cs_2AgBiBr_6$ with the available experimental data for all-inorganic halide perovskites at 300 K [12-15,72,73], as illustrated in Fig. 3(d). Our predicted total $\kappa_L$ for $Cs_2AgBiBr_6$ is within the same order of magnitude ($\leq 0.64$ Wm$^{-1}$K$^{-1}$) as the $\kappa_L$ values reported for other halide perovskites [see Fig. 3(d)]. After a thorough examination, we discovered that the experimental $\kappa_L$ of halide perovskites tends to decrease as the structural complexity increase. Structural complexity is typically characterized by the size of the unit-cell and the types of elements present in the halide perovskite crystal. An experiment recently reported an ultralow room-temperature $\kappa_L$ of 0.31 Wm$^{-1}$K$^{-1}$ in crystalline $Cs_2SnI_6$ [72]. Given the similarity in structure and phonon dispersions between $Cs_2SnI_6$ and $Cs_2AgBiBr_6$ [75], we can expect a similar result for $Cs_2AgBiBr_6$. Furthermore, the frequency of soft mode at Γ point in $Cs_2SnI_6$ is around 0.6 THz, which is higher than that in $Cs_2AgBiBr_6$ (~0.4 THz) at 300 K. Therefore, we can expect a lower experimental $\kappa_L$ in $Cs_2AgBiBr_6$ as compared to $Cs_2SnI_6$. An even lower $\kappa_L$ value of 0.19 Wm$^{-1}$K$^{-1}$ was observed for the halide perovskite $Cs_3Bi_2I_6Cl_3$, which has a higher structural complexity than $Cs_2AgBiBr_6$, in a recent experiment at room temperature [15]. Therefore, the $\kappa_L$ of $Cs_2AgBiBr_6$ is expected to be higher than 0.19 Wm$^{-1}$K$^{-1}$. Overall, the predicted total $\kappa_L$ of $Cs_2AgBiBr_6$ in this study is considered to be reasonably reliable and can be verified by future experimental investigations.

### c) Microscopic mechanisms of populations' conductivity

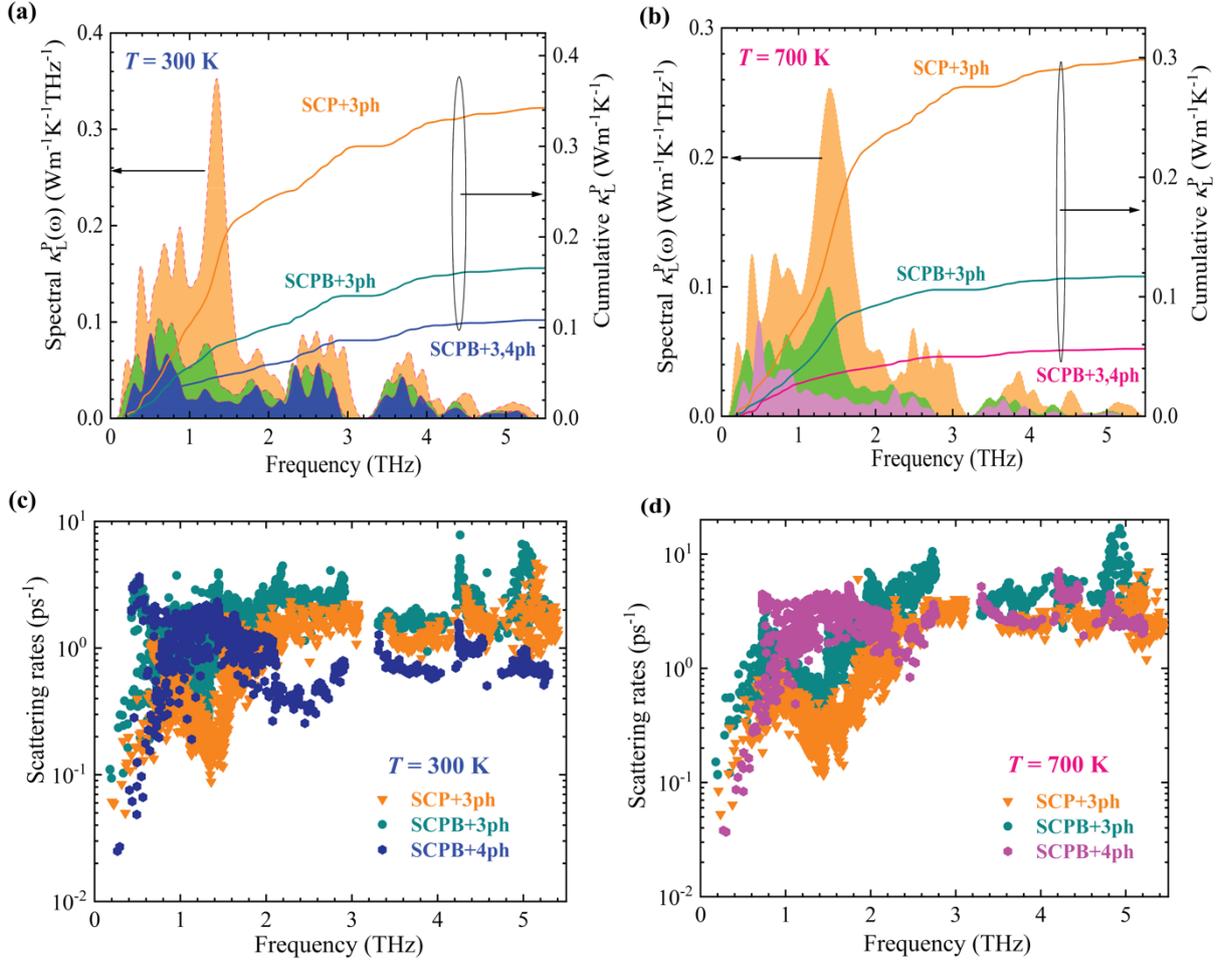

FIG. 4. Comparison of spectral/cumulative phonon thermal conductivity and scattering rates calculated using different levels of theory, i.e., SCP+3ph and SCPB+3/3,4ph models, at $T = 300$ and $700$ K, respectively. (a) Spectral and cumulative phonon thermal conductivities at $T = 300$ K. (b) The same as (a), but $T = 700$ K. (c) 3ph and 4ph scattering rates at $T = 300$ K. (d) The same as (c), but $T = 700$ K.

To better illustrate the effect of anharmonic phonon renormalization and higher-order phonon scattering processes on the $\kappa_L^P$ of $Cs_2AgBiBr_6$, we present the spectral and cumulative $\kappa_L$ in Figs. 4(a-b) calculated using SCP+3ph and SCPB+3/3,4ph models at 300 and 700 K, respectively. For all the three levels of theory, our results consistently show that most of populations' thermal conductivity $\kappa_L^P$ in $Cs_2AgBiBr_6$ is carried by phonons with a frequency less than 3 THz across the temperature range of $300 - 700$ K [see Figs. 4(a-b)]. Additionally, we observed that optical phonons with a frequency higher than 1.2 THz play a dominant role in the $\kappa_L^P$ from the particle-like propagation channel. This behaviour is commonly observed in cubic SrTiO$_3$ with strong anharmonicity and low-$\kappa_L^P$ [22], and is attributed to the suppression of acoustic phonons by highly anharmonic optical modes.

The spectral and cumulative $\kappa_L^P$ obtained from the SCP+3ph model show a reduction compared to the results from the SCPB+3ph model. This decrease in $\kappa_L^P$ is mainly contributed by the phonons with frequency less than 3 THz [see Figs. 4(a-b)]. This is because the cubic anharmonicity contributes negative phonon energy shifts, which results in the softening of phonons and enhancement of phonon scattering processes [see Figs. 4(c-d)]. With the additional 4ph scattering processes, the reduced $\kappa_L^P$ obtained by SCPB+3,4ph relative to SCPB+3ph is primarily contributed by phonons with a frequency less than 3 THz at both 300 and 700 K. This can be attributed to the inherently strong 4ph scattering rates of specific modes, such as rattling-like flat modes, which will be demonstrated later.

By examining the group velocity, specific heat, and scattering rates, in computing $\kappa_L^P$ [34], the reduction in $\kappa_L^P$ in two improved models (SCPB+3/3,4ph) relative to the SCP+3ph model is mainly attributed to the scattering rates [see Figs. 4(c-d)]. This is due to the fact that the difference in group velocity and specific heat between the SCP and SCPB approximation is negligible in Cs$_2$AgBiBr$_6$, as demonstrated by see Figs. S7 and S8 in SM [44]. Indeed, previous studies have shown that the group velocity and specific heat of certain crystals are relatively insensitive to phonon energy shifts resulting from lattice anharmonicity [25,54-56]. As shown in Figs. 3(b) and 4(a-b), the inclusion of additional 4ph scatterings in the calculation of $\kappa_L^P$ (using the SCPB+3,4ph model) leads to a significant reduction compared to considering only 3ph processes (using the SCPB+3ph model). Specifically, the reduction in $\kappa_L^P$ is 34.6 and 51.8% at 300 and 700 K, respectively. This can be ascribed to the large 4ph scattering rates, which are roughly as high as those of 3ph processes at 300 K, and even become dominant at 700 K, particularly for phonons with frequency less than 2 THz, as shown in Figs. 4(c-d).

It is worth noting that both the 3ph and 4ph scattering rates decrease with increasing frequency for phonons below 1 THz ( shown in the upper left part of Figs. 4(c-d)). Upon analysis of the distribution of scattering rates and atomic participation ratios [see Figs. 4(a-b) and Figs. S2 in SM [44]], we have found that the observed phenomenon can be attributed to the presence of Br-dominated soft modes, and a similar phenomenon was also reported for other compounds, e.g., SrTiO3, PbSe and PbTe [22,76]. Generally, the ultra-low $\kappa_L$ observed in in halide perovskites is attributed to the combination of strong scattering rates and small group velocity [19,77]. The small group velocity with a maximum value of ~2.35 Km/s observed in Cs$_2$AgBiBr$_6$ may lead to the ultra-low $\kappa_L$ [see Fig. S7 in SM [44]]. However, neglecting the contribution from Br-dominated soft modes in 3ph scattering processes calculation leads to a significant enhancement of $\kappa_L^P$, with a value of ~87 Wm$^{-1}$K$^{-1}$ at 300 K [see Figs. S9(a) and S10 in SM [44]]. Similarly, neglecting Cs-dominated modes leads to an increase in $\kappa_L^P$ to 18 from ~0.17 Wm$^{-1}$K$^{-1}$ calculated using SCPB+3ph [see Figs. S9(b) and S10 in SM [44]]. This illustrates the important role of the Br-dominated soft modes with strong anharmonicity [see Fig. S9(a), S11 and S12 in SM [44]] in contributing to the ultra-low $\kappa_L^P$ of Cs$_2$AgBiBr$_6$. Overall, the strong anharmonicity and resulting high scattering rates are the dominant factors in the ultra-low $\kappa_L$ of Cs$_2$AgBiBr$_6$.

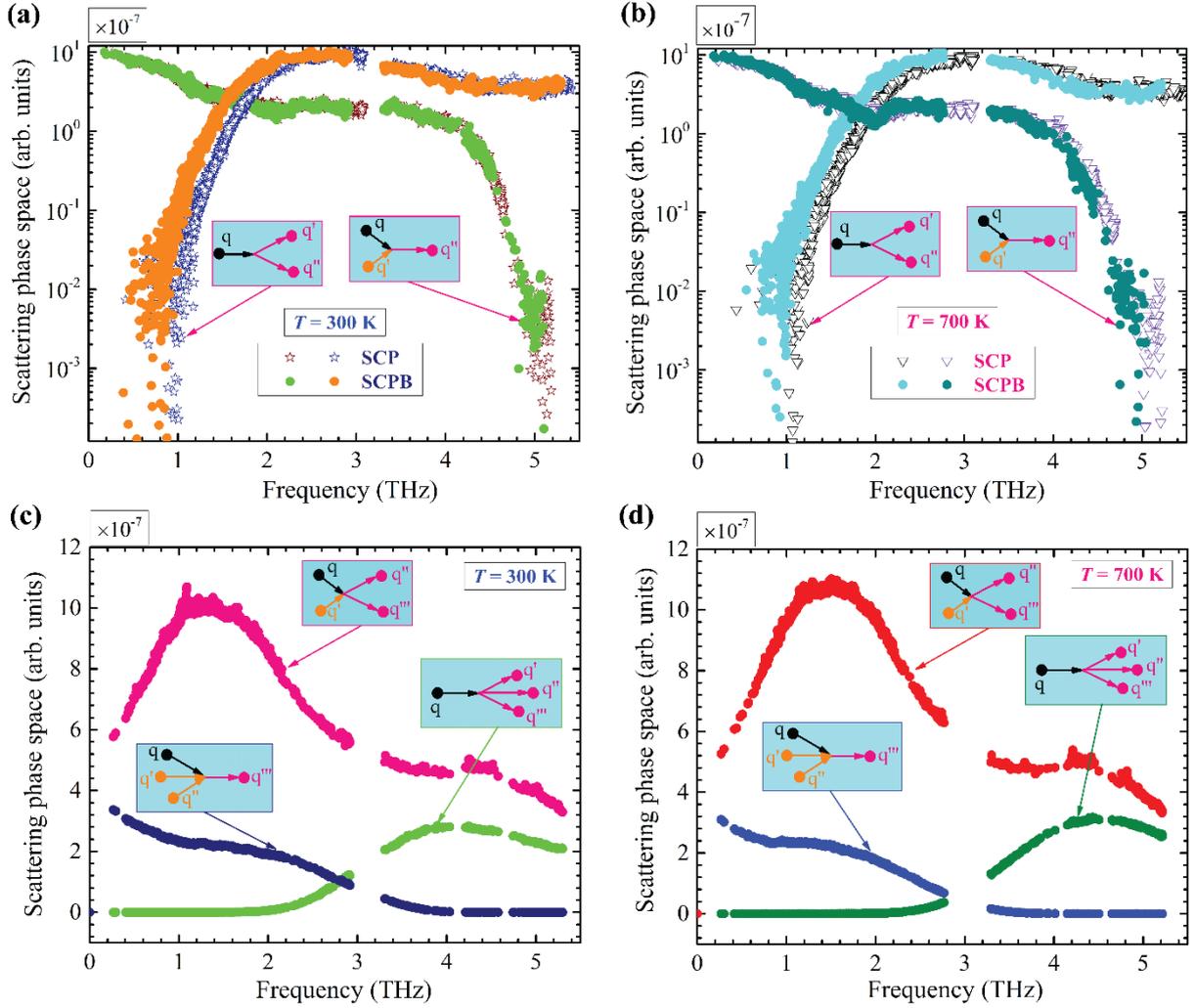

FIG. 5. (a) Comparison of mode-resolved 3ph scattering phase space calculated using SCP and SCPB approximations at $T = 300$ K, respectively. The 3ph scattering phase spaces are decomposed to the absorption ($q + q' \rightarrow q''$) and emission ($q \rightarrow q' + q''$) processes, respectively. (b) The same as (a), but at $T = 700$ K. (c) Mode-resolved 4ph scattering phase space calculated using SCPB approximation at $T = 300$ K, wherein the 4ph scattering phase spaces are decomposed to the recombination ($q + q' + q'' \rightarrow q'''$), redistribution ($q + q' \rightarrow q'' + q'''$) and splitting ($q \rightarrow q' + q'' + q'''$) processes, respectively. (d) The same as (c), but at $T = 700$ K.

To reveal the correlation between the negative phonon energy shift due to cubic anharmonicity and the resulting change in phonon scattering rates, we compare scattering phase space [59,63] calculated using the SCP and SCPB approximation at 300 and 700 K, as shown in Figs. 5(a-b). The scattering phase space of emission processes computed with SCPB+3ph is substantially larger relative to the SCP+3ph results for phonons with a frequency less than 3 THz at 300 and 700 K. This observation suggests that the negative phonon shifts resulting from cubic anharmonicity leads to enhance the scattering phase space and thus stronger phonon scattering rates. Similar observations have been also reported in other crystals such as $BaZrO_3$ and CsPbBr$_3$ [55,57]. Similarly, the enhanced scattering rates for phonons with frequencies greater than 3 THz between the SCP+3ph and SCPB+3ph models can be attributed to the enhanced scattering phase space of absorption processes. Note that another key quantity in calculating scattering rates, namely, the scattering matrix element, is insensitive to the phonon energy shift due to lattice anharmonicity [54].

To gain a deeper insight into the microscopic origin of strong 4ph scattering rates in Cs$_2$AgBiBr$_6$, we calculated the scattering phase space for 4ph processes at 300 and 700 K, respectively, as shown in Figs. 5(c-d). We found that the scattering phase space for 4ph processes, specifically for the redistribution processes ($q + q' \rightarrow q'' + q'''$), is remarkably strong in Cs$_2$AgBiBr$_6$, with magnitudes comparable to those of 3ph processes at 300 K and even surpassing them at 700 K [see Figs. 5(a-d)]. This observation is in line with the magnitude distribution between the 3ph and 4ph scattering rates [see Figs. 4(c-d)] and provides further evidence that the large 4ph scattering phase space contributes to strong 4ph scattering rates in Cs$_2$AgBiBr$_6$ [see Figs. 5(c-d)]. The peak region of the 4ph redistribution processes in Cs$_2$AgBiBr$_6$ was found to coincide with the partial DOS of Cs atoms [see Figs. 1(c) and 5(c-d)], which suggests that Cs atoms play a significant role in the strong 4ph scattering rates. Additionally, the atom participation ratio calculated in Fig. S13 of SM [44] reveals that Cs atoms contribute to flat phonon bands and exhibit rattling-like behaviours (This observation can be attributed to the loose bonding of Cs atoms [see Fig. S14 of SM [44]]), which are known to result in strong scattering phase space and scattering rates [56,78]. Flat rattling-like modes contributing to strong 4ph scatterings have also been observed in other materials such as crystalline AgCrSe$_2$ [79] and Cu$_{12}$Sb$_4$S$_{13}$ Tetrahedrites [56].

### d) Microscopic mechanisms of coherences' conductivity

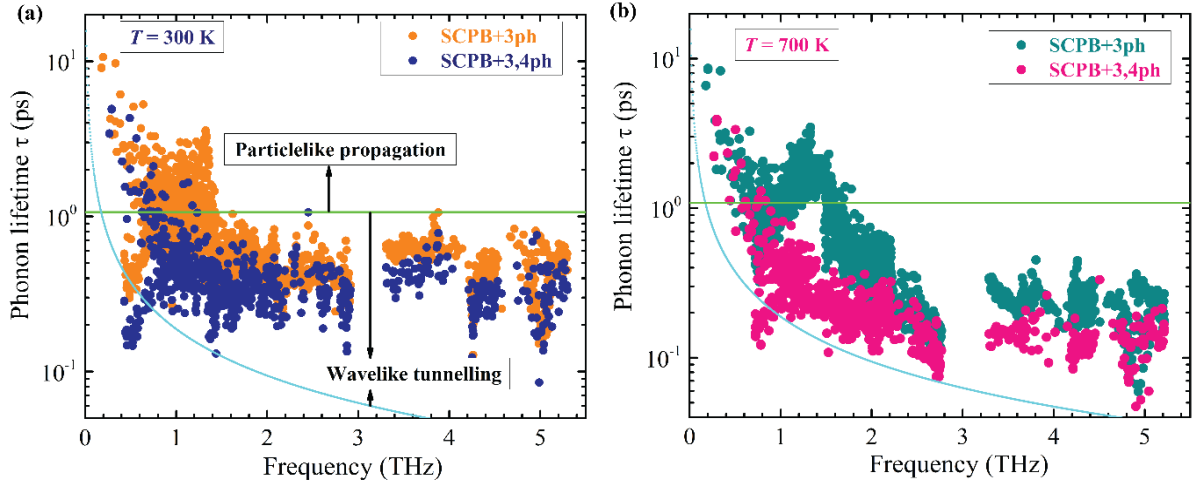

FIG. 6. (a) Calculated phonon lifetimes as a function of frequency at $T$ = 300 K within SCPB+3/3,4ph models, respectively. The green solid line depicts the Wigner limit in time [61] and can be written as $\tau = [\Delta\omega_{avg}]^{-1}$, where $\Delta\omega_{avg}$ is the average phonon inter-band spacing and can be defined as $\omega_{max}/3N_{at}$, where $\omega_{max}$ denotes the highest frequency and $N_{at}$ denotes the number of atoms in primitive cell. Phonons with lifetime larger than this limit mainly contribute to populations' thermal conductivity; on the contrary, phonons mainly contribute to coherences' thermal conductivity. The Cyan dots indicates the Ioffe-Regel limit in time [80], i.e., $\tau = 2\pi/\omega$, $\omega$ is the phonon frequency with unit of cm$^{-1}$. Phonons with lifetime larger than the Ioffe-Regel limit in time are recognized as well-defined and can be handled by the unified theory of thermal transport [30]. (b) The same as (a), but $T$ = 700 K.

As shown in Figs. 3(b-c), the dominant contribution to the total $\kappa_L$ above ~340K is from the coherences' contributions. Specifically, using the SCPB+3,4ph+OD model, the coherences' conductivity $\kappa_L^C$ accounts for 48.4 and 60.3% of the total $\kappa_L$ at 300 and 700 K, respectively. To uncover the microscopic origin of wavelike conduction in crystalline Cs$_2$AgBiBr$_6$, we calculated the phonon lifetime as a function of frequency from 3ph and 4ph processes at 300 and 700 K, respectively, as shown in Figs. 6(a-b). Recently, a new concept called the Wigner limit in time proposed by Simoncelli *et al.* [61] was used to separate phonons into different thermal transport regimes, mainly including particle-like propagation and wave-like tunnelling regimes. The Wigner limit in time is defined as $\tau = 3N_{at}/\omega_{max}$, where $N_{at}$ is the number of atoms in the primitive cell and $\omega_{max}$ is the maximum phonon frequency. Using the criterion of Wigner limit in time, phonons with a lifetime above this limit behave in a particle-like propagation and mainly contribute to the populations' thermal conductivity; on the contrary, phonons behave in a wave-like tunnelling and mainly contribute to the coherences' thermal conductivity.

For the transport regimes in crystalline $Cs_2AgBiBr_6$, after considering both 3ph and 4ph scattering processes, most phonons have a lifetime below the Wigner limit in time at both 300 and 700 K, highlighting the necessity of evoking wavelike tunnelling transport channel [see Fig. 6(a-b)]. These findings suggest that the coherences' conductivity from wavelike tunnelling channel may dominate over the total thermal conductivity in $Cs_2AgBiBr_6$, consistent with the results shown in Figs. 3(b-c). When only 3ph scattering processes are considered, it is shown in figures 6 (a) and (b) that there are many phonons with a lifetime above the Wigner limit in time. This suggests that the wave-like tunnelling channel may be as important as the particle-like propagation channel in $Cs_2AgBiBr_6$ [see Figs. 3(c) and S15 in SM [44]]. Interestingly, compared with the $\kappa_L^C$ calculated by the SCPB+3ph+OD model, further including 4ph scattering processes, namely, SCPB+3,4ph+OD, significantly suppresses the particle-like propagation channel. As a result, the wave-like tunnelling channel plays a dominant role in thermal transport in $Cs_2AgBiBr_6$ [see Fig. 3(c)]. This indicates the conversion of dominant thermal transport channels between the particle-like phonon propagation and wave-like tunnelling channels. These findings emphasize the inadequacy of the conventional phonon-gas model in describing thermal transport in $Cs_2AgBiBr_6$. They also highlight the crucial role of comprehensive modelling, such as the SCPB+3,4ph+OD model, in gaining a deeper understanding of the actual heat transfer mechanism in $Cs_2AgBiBr_6$.

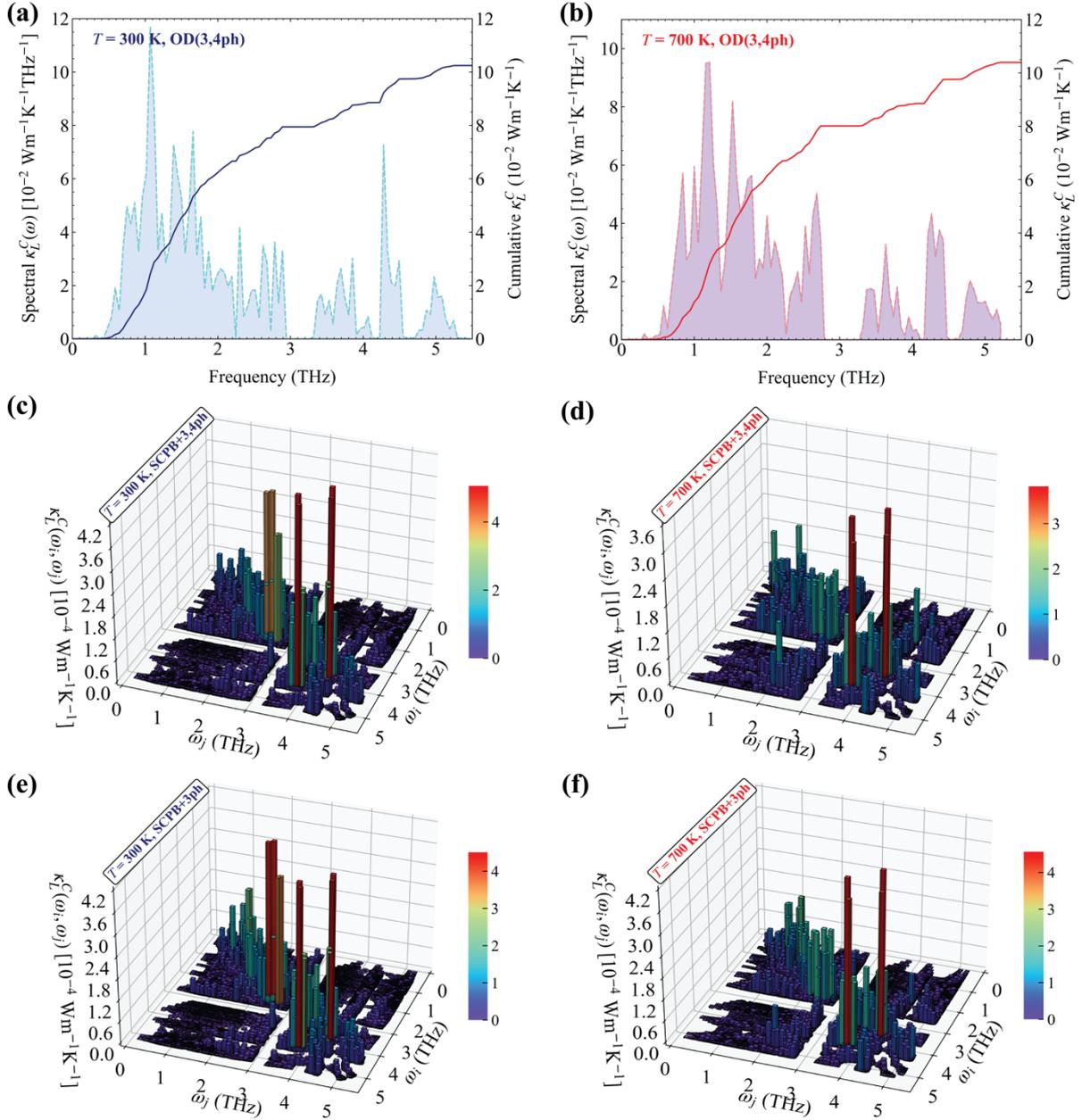

FIG. 7. (a) Calculated spectral and cumulative coherences' thermal conductivity using SCPB+3,4ph+OD model at $T = 300$ K. The contribution of phonon mode $j$ in coherence couplings between two phonon modes $(\mathbf{q}, j)$ and $(\mathbf{q}, j')$ is evaluated by $c_{\mathbf{q}j}/(c_{\mathbf{q}j} + c_{\mathbf{q}j'})$, where $c_{\mathbf{q}j}$ is the phonon mode-specific heat. (b) The same as (a) but at 700 K. (c) Two-dimensional (2D) modal $\kappa_L^C(\omega_{\mathbf{q}j}, \omega_{\mathbf{q}j'})$ of the contributions to the coherences' thermal conductivity calculated using the SCPB+3,4ph+OD model at $T = 300$ K. The diagonal data points ($\omega_{\mathbf{q}j} = \omega_{\mathbf{q}j'}$) indicate phonon degenerate eigenstates. (d). The same as (c) but $T = 700$ K. (e) The same as (c) but using SCPB+3ph+OD model. (f) The same as (e) but at $T = 700$ K.

To gain a better understanding of microscopic mechanisms behind the coherences' thermal conductivity $\kappa_L^C$ in Cs$_2$AgBiBr$_6$, we calculated the 1D-spectral and cumulative mode-specific contributions to coherences' conductivity at 300 and 700 K, respectively, as shown in Figs. 7(a-b). Similar to the spectral populations' conductivity $\kappa_L^P$ of Cs$_2$AgBiBr$_6$ in Figs. 4(a-b), the majority of the coherences' conductivity from the wavelike tunnelling channel is carried by the phonons with a frequency less than 3 THz at both 300 and 700 K [see Figs. 7(a-b)]. This can be attributed to the small inter-band spacing (due to dense phonon dispersions [see Fig. 1(b)]) in conjunction with large linewidths (large scattering rates [see Figs. 4(a-b)] or strong anharmonicity [30]) within frequency region below 3 THz [61]. This phenomenon is consistent with the observation presented in Figs. 6(a-b).

Figures 7(c-d) show that the contributions to coherences' thermal conductivity in $Cs_2AgBiBr_6$, calculated using SCPB+3,4ph+OD at 300 and 700 K, can be resolved in terms of the phonon energies, namely, $\omega_{qj}$ and $\omega_{qj'}$, of two coupled phonons. In contrast to the $\kappa_L^C$ in $CsPbBr_3$ at 50 K [30] and $La_2Zr_2O_7$ at 200 K [61], where quasi-degenerate phonon states dominate, the phonons contributing to the $\kappa_L^C$ of $Cs_2AgBiBr_6$ at both 300 and 700 K have relatively different frequencies [see Figs. 7(c-d)]. This suggests that the contributions to the $\kappa_L^C$ are driven by the strong anharmonicity and the corresponding heat transfer physics of wave-like coherences in $Cs_2AgBiBr_6$ is intrinsically different from that of harmonic glasses. By comparing the contributions to the $\kappa_L^C$ in Figs. 7(c) and (d), it is observed that couplings between phonons with large frequency differences begin to contribute to the $\kappa_L^C$ with increasing temperature [for instance, when considering coupling pairs with a frequency difference larger than 2 THz, their coherences' contribution $\kappa_L^C$ increases by ~ 6.9% with the temperature rising from 300 to 700 K]. This is because the scattering rates of phonons, i.e., their linewidths, also increase as the temperature increases. Additionally, a comparison between the contributions to the $\kappa_L^C$ calculated using SCPB+3ph+OD and SCPB+3,4ph+OD models [see Figs. 7(c-f)] reveals that the additional 4ph scattering rates result in more scattered contributions to $\kappa_L^C$. This is because the large 4ph scattering rates allow phonons with relatively large frequency differences to couple, particularly those involving the Cs rattling-like modes. This phenomenon is similar to what was observed in $Cu_{12}Sb_4S_{13}$ Tetrahedrites [56].

To this end, the application of the SCPB+3,4ph+OD model has provided a reliable prediction of the $\kappa_L$ in crystalline $Cs_2AgBiBr_6$, and has elucidated its microscopic mechanism of heat transport within the temperature range of 300 - 700 K. However, further improvement may be possible by accounting for (i) the anharmonic contributions of heat flux operators [71], (ii) simultaneously evaluating phonon energy shifts and broadening.

## IV. CONCLUSIONS

In conclusion, we have performed the DFT calculations to investigate the anharmonic lattice dynamics and unravel the microscopic mechanisms of thermal transport in crystalline $Cs_2AgBiBr_6$. Our results show that the unstable soft modes correspond to the in-phase and out-of-phase tilting of the $AgBr_6$ and $BiBr_6$ octahedra units in real space. The anharmonic stabilization of the soft modes is observed to occur at ~119-138 K in excellent agreement with experiments. We found that the negative phonon energy shifts arising from cubic anharmonicity plays a crucial role in accurately reproducing experimental phonon dispersions. The accurate reproduction of experimental phonon energies by the SCPB approximation underscores the significance of considering both cubic and quartic anharmonicities in anharmonic phonon renormalization.

With advanced unified theory of thermal transport, the $\kappa_L$ of $Cs_2AgBiBr_6$ is predicted to be ultra-low, with a value of 0.21 $Wm^{-1}K^{-1}$ at 300 K, which can be mainly attributed to the inherently strong scatterings rather than small group velocities. The negative phonon energy shifts from cubic anharmonicity leads to enhanced phonon scattering rates, thus reducing $\kappa_L$ in $Cs_2AgBiBr_6$. Additionally, our results indicate that the 4ph scatterings significantly suppress the $\kappa_L$ in $Cs_2AgBiBr_6$. Through analysis of the phonon scattering phase space for 3ph and 4ph processes, we attribute the strong 4ph scattering rates to the flat phonon modes dominated by Cs atoms.

It is also revealed that the coherences' conductivity dominates the total $\kappa_L$ above 310 K, highlighting the breakdown of the conventional phonon gas model in accurately predicting the $\kappa_L$ of $Cs_2AgBiBr_6$. The comparison of the results obtained from SCPB+3ph+OD and SCPB+3,4ph+OD models shows that the inclusion of 4ph scattering processes alters the dominance of wave-like tunnelling and particle-like phonon propagation channels in thermal transport of $Cs_2AgBiBr_6$. Thus, both cubic and quartic anharmonicities are critical in determining not only the anharmonic lattice dynamics but also lattice thermal conductivity. Our study provides insights into the microscopic mechanisms of thermal transport in halide double perovskites with ultra-low $\kappa_L$ and strong anharmonicity.

## ACKNOWLEDGEMENT

We are thankful for the financial support from the Science and Technology Planning Project of Guangdong Province, China (Grant No. 2017A050506053), the Science and Technology Program of Guangzhou (No. 201704030107), and the Hong Kong General Research Fund (Grants No. 16214217 and No. 16206020). This paper was supported in part by the Project of Hetao Shenzhen-Hong Kong Science and Technology Innovation Cooperation Zone (HZQB-KCZYB2020083). R.G. acknowledges support from the Excellent Young Scientists Fund (Overseas) of Shandong Province (2022HWYQ091) and the Initiative Research Fund of Shandong Institute of Advanced Technology (2020107R03). G.H. and J.Z. acknowledge funding by the U.S. Department of Energy, Office of Science, Office of Basic Energy Sciences, Materials Sciences and Engineering Division, under Contract No. DE-AC02-05-CH11231:

Materials Project program KC23MP. C.L. acknowledges the support from the Sinergia project of the Swiss National Science Foundation (grant number CRSII5_189924).

There are no conflicts of interest to declare.